# Strain engineering of the magnetic anisotropy and moment in $NdFeO_3$ epitaxial thin films

M. A. Khaled [1], J. Ruvalcaba [2], T. Cordova-Fraga [2], D. C. Arnold [3], N Jaouen [4], P Ohresser [4], M. Jouiad [1], K. Hoummada [5], B. Dkhil [6], M. El Marssi [1], H. Bouyanfif [1,*]

[1] *LPMC UR2081, Université de Picardie Jules Verne, 33 Rue Saint Leu, 80000 Amiens, France*

[2] *División de Ciencias e Ingenierías, Universidad de Guanajuato campus León, México*

[3] *School of Physical Sciences, University of Kent, Canterbury, Kent, CT2 7NH, UK*

[4] *Synchrotron SOLEIL, L'Orme des Merisiers, Saint-Aubin, BP 48, 91192 Gif-sur-Yvette Cedex, France*

[5] *Aix Marseille Univ, CNRS, Université de Toulon, IM2NP, 13397, Marseille, France*

[6] *Université Paris-Saclay, CentraleSupélec, Laboratoire Structures, Propriétés et Modélisation des Solides (SPMS), CNRS UMR8580, 91190 Gif-sur-Yvette, France*

**Abstract:** Strain engineering is a powerful mean for tuning the various functionalities of $ABO_3$ perovskite oxide thin films. Rare-earth orthoferrite $RFeO_3$ materials such as $NdFeO_3$ (NFO) are of prime interest because of their intriguing magnetic properties as well as their technological potential applications especially as thin films. Here, using a large set of complementary and advanced techniques, we show that NFO epitaxial thin films, successfully grown by pulsed laser deposition on (001)-$SrTiO_3$, show a strong magnetic anisotropy below a critical thickness $t_c$ of ~54 nm, associated to the occurrence of structural modifications related to symmetry and domain pattern changes. By varying the tensile misfit strain through the decrease of film thickness below $t_c$, the amplitudes of in- and out-of-plane magnetization can be continuously tuned while their ratio keeps constant. Furthermore, different low temperature magnetic behaviours are evidenced for strained and relaxed films suggesting that the strain-induced structural state impacts the magnetic phase stability.

## I. Introduction

Orthoferrites, $RFeO_3$ ($R^{3+}$: rare earth) have received a lot of attention due to their fascinating magnetic properties originating from the two magnetic ions (i.e. the $R^{3+}$ rare earth and $Fe^{3+}$ transition metal ions) present in different sublattices and their cross-talk interactions ($Fe^{3+}$-$Fe^{3+}$, $Fe^{3+}$-$R^{3+}$, $R^{3+}$-$R^{3+}$). Bulk $RFeO_3$ materials crystallize in an orthorhombic structure with a canted antiferromagnetic (AFM) spin ordering leading to a non-collinear magnetism through Dzyaloshinskii-Moriya interaction (DMI) [1-3]. Their high Néel temperature ensuring a robust magnetism (AFM) against external magnetic perturbation makes them of interest for applications such as terahertz emitters (ultrafast spin dynamic) and data storage and processing devices [4-7]. Among $RFeO_3$, Neodymium ferrite $NdFeO_3$ (NFO) is an antiferromagnet that has been often used as a prototypical model for experimental and computational investigations [8]. The perovskite bulk structure adopts a distorted orthorhombic Pbnm phase with lattice parameters $a_O$ = 5.451 Å, $b_O$ = 5.588 Å, $c_O$ = 7.761 Å at room temperature (RT) [9]. The orthorhombic unit cell being doubled along the $c_O$ axis and rotated by 45° in the ($a_O$, $b_O$) plane relative to the pseudo-cubic (pc) unit cell ($a_{pc}$ ~ 3.9 Å and see Fig S1 for a sketch of the bulk NFO unit cell and spin structure and the possible domains on a cubic substrate) [10]. NFO magnetic properties are mainly attributed to the superexchange interaction via Fe-O-Fe bonds inducing a canted G-type AFM ordering with a Néel temperature of $T_N$ ($Fe^{3+}$) ≃ 760 K [1,11]. A spin reorientation transition (SRT) has been observed for bulk NFO and shown to occur in the temperature range of 100 K to 170 K [8,12]. At $T_{SRT}$, the $Fe^{3+}$ magnetic structure in NFO undergoes a spin reorientation transition from a {Gx, Mz} ordering at high temperature to a {Gz, Mx} ordering at lower temperature. Here, {$G_x$, $M_z$} (respectively {$G_z$, $M_x$}) means G-type AFM ordering of $Fe^{3+}$ spins with their moment direction along the "a" (respectively "c") orthorhombic axis which is the easy anisotropy direction and $M_z$ (respectively $M_x$) is the weak ferromagnetism along "c" axis (respectively "a" axis) resulting from the $Fe^{3+}$ spin canting due to DMI. In contrast to other nonmagnetic $R^{3+}$ elements (La, Eu, Lu), NFO exhibits a second magnetic sublattice due to partially occupied 4f orbitals of $Nd^{3+}$. The $Nd^{3+}$ spin contribution becomes significant at low temperature and its antiferromagnetic-like interactions with $Fe^{3+}$ spins led to a spontaneous spin reversal of the two Fe and Nd sublattices as well as magnetic moment compensation because the effective moment of the Nd sublattice increases faster than that of the antiparallel Fe sublattice when the temperature decreases, at $T_{SR}$ ~ 29 K and $T_{comp}$ ~ 8 K respectively [8,12]. It was also revealed that the two distinct sublattices sustain a strong interaction accompanied by a spin-lattice coupling. As a matter of fact, the magnetic background instabilities of NFO result in intriguing phenomena such as magnetization reversal, large magnetic anisotropy, and spin switching [8]. For obvious reasons of miniaturization and applications, orthoferrites thin films have been studied for several magnetic rare earth elements such as Sm, Gd, Dy, and Tm. High-quality samples have been synthesized on perovskite oxide substrates ($SrTiO_3$, $LaAlO_3$) using different deposition techniques such as pulsed laser deposition (PLD) or magnetron sputtering [5,13-15]. Most of the films exhibit a Fe sublattice bulk-like antiferromagnetism at high temperatures with a SRT upon cooling. Structural studies of $RFeO_3$ show an orthorhombic structure with a complex twinning pattern depending on the mismatch and strain relaxation associated with the substrate. Orthoferrites are very sensitive to strain effects and for instance, SRT can be manipulated by means of epitaxial strain as reported for $DyFeO_3$ grown on $SrTiO_3$ (001) substrates (imposing compressive strain) [13]. $SmFeO_3$ thin films grown on $SrTiO_3$ (001) substrates were also investigated and modulation of the magnetic properties with thickness is evidenced [14]. More recently, S. Becker *et al.* demonstrated that the SRT readout could be electrically achieved in $TmFeO_3$ thin films by making use of the spin Hall magnetoresistance [5] opening an avenue for electric-manipulation of the spin moment in orthoferrites. Interestingly, recent theoretical studies have also demonstrated additional phenomena in $NdFeO_3$-based superlattices such as improper ferroelectricity and exotic magnetoelectric coupling [16] widening the potentialities of orthoferrites as nanostructures. Despite of the flurry of phenomena in orthoferrites to be investigated and explored, to the best of our knowledge no experimental work on epitaxial NFO thin films has been reported yet, probably due to generic issues of secondary phase formation (garnet, iron oxide) during growth process. Here, we successfully grow on $SrTiO_3$ (001)-oriented substrate parasitic-phase free epitaxial NFO films with thickness varying between 5 nm and 212 nm. Using complementary and advanced electron microscopy, synchrotron X-ray absorption spectroscopy, vibrating sample magnetometry and high-resolution diffraction and reflectivity techniques, we provide a full structural investigation and characterization of the magnetic behaviour showing that thickness engineering enabling to tune the misfit strain is an efficient leverage for fully controlling the magnetic anisotropy and amplitude of the magnetic moment of NFO films.

## II Experimental details

The NFO thin films (thickness ranging from 5 nm to 212 nm) were grown on $SrTiO_3$ (001) oriented substrates (STO) using pulsed laser deposition (PLD) equipped with an excimer KrF laser (248 nm wavelength). A homemade NFO ceramic target was used for the thin film growth (supplementary information) [10]. Reflection high energy electron diffraction (RHEED) allowed us to in-situ check the surface quality and growth of the NFO thin film (supplementary information figure S2) [10]. The optimized thin film deposition parameters are presented in table S1. High-resolution X-Ray diffraction (XRD) characterizations (θ-2θ, rocking curve, reflectivity, φ-scan, reciprocal space mapping) were performed using a D8 Brucker diffractometer ($\lambda$ = 1.54056 Å). Transmission electron microscopy (TEM) cross-sectional analysis was performed on lamella prepared by Focused Gallium Ion Beam (FIB). The obtained lamella thicknesses were about 100 nm by milling using a Thermo Fisher dual-beam HELIOS 600 nanolab setup. HRTEM images were collected with a Field Emission Gun (FEG) Titan microscope by FEI operated at 200 kV. The microscope was equipped with a spherical aberration correction (Cs) system and a contrast diaphragm of 60 μm. Synchrotron X-ray absorption spectroscopy was performed at SOLEIL synchrotron at DEIMOS beamline with the incident beam normal to the sample surface with vertical and horizontal linear X-ray polarizations defined in the laboratory frame (LH: linear horizontal; LV: Linear vertical) [17]. A Quantum Design Cryogen-free system equipped with a Vibrating Sample Magnetometer (VSM) allowed us to acquire magnetic moment evolution with temperature and M-H hysteresis loops from 3 K to 300 K.

## III Results and discussion

Figure 1 presents the room temperature X-ray diffraction of the whole set of samples. In order to check the interface quality and thickness of the samples, X-ray reflectivity scans were collected and are presented in figure 1 (a). The large number of oscillations obtained on the reflectivity scans is a proof of the flat nature and high quality of the interfaces. The simulations of the reflectivity (inset figure 1 (a)) enable us to verify the thickness of the NFO thin films for a certain range of thicknesses (up to 81 nm thickness). The NFO thin films are fully epitaxial with (00l) pseudo-cubic orientation along the growth direction and free of parasitic phases for the whole set of samples (see θ-2θ figure 1 (b)). Zooming in on the angular range of the first order of diffraction of the STO for the thinnest samples (figure 1 (c) and S3) [10] further attests to the very high quality of the thin NFO layers. Indeed, Laue oscillations (figure 1 (c) and S2) [10] are observed and are signatures of very high crystalline quality, sharp interfaces, and very flat top surfaces. Simulations of the Laue oscillations confirm the thickness of the samples obtained by reflectivity (see Figure S3) [10].

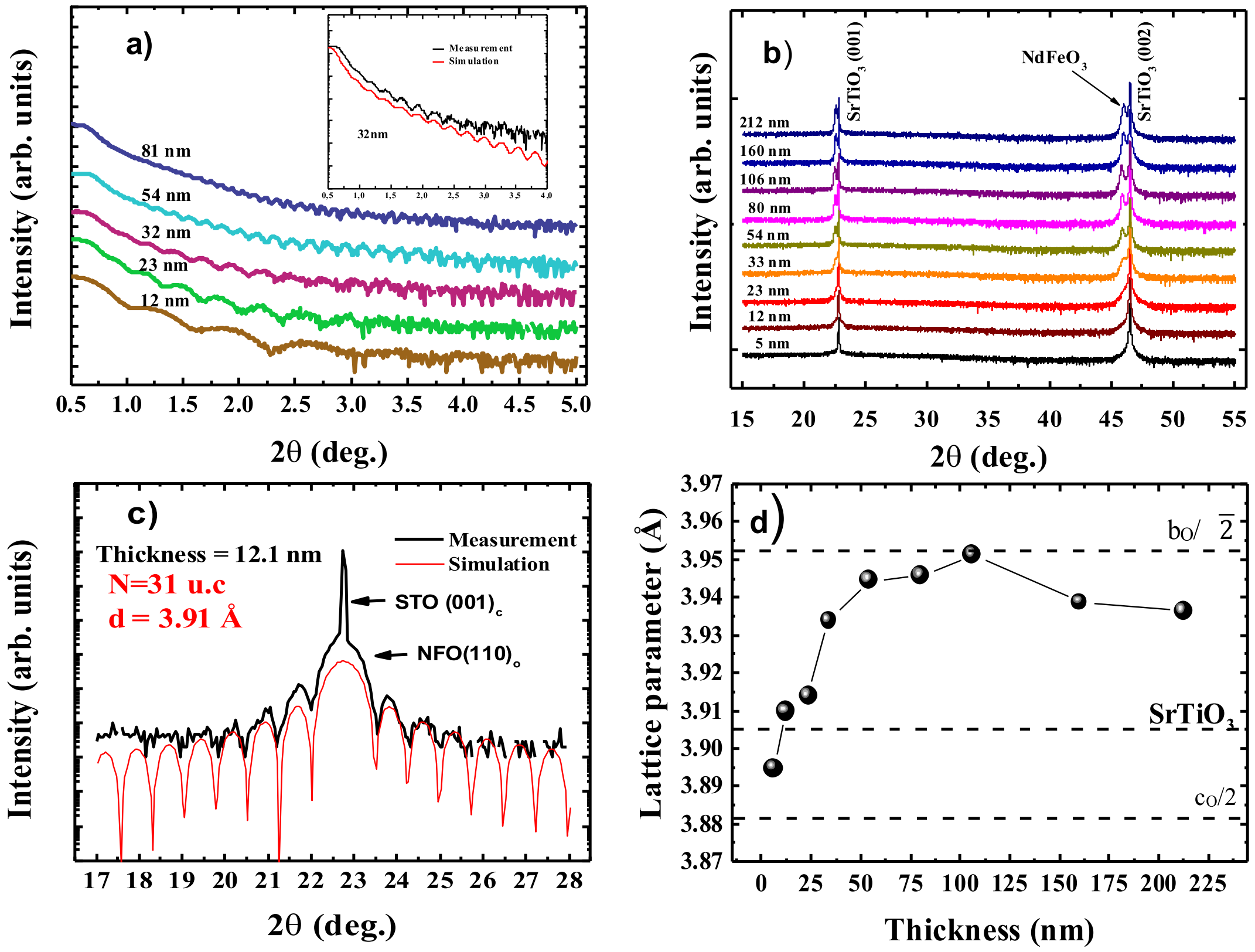

Figure 1. X-ray diffraction analysis of the NFO thin films. (a) Reflectivity scans for thin films up to 81 nm thickness. The inset shows a representative simulation for such a reflectivity scan. (b) θ-2θ diffraction pattern for the whole set of NFO thin films. Note the NFO peaks shift to higher angles on increasing thickness. (c) θ-2θ diffraction pattern around the 1$^{st}$ order peak of the STO substrate (intense and sharp peak) for highlighting the presence of Laue oscillations (see text). (d) Evolution of the NFO pseudo-cubic out-of-plane lattice parameter versus thickness. The dashed horizontal line indicates the lattice parameter of the STO substrate.

Figure 1 (b) also sheds light on a structural trend of the NFO thin films on tuning the thickness. A Bragg peak shift to a high angle for NFO is evidenced by decreasing the thickness. This peak shift corresponds to a decrease of the out-of-plane lattice parameter on thickness decreasing, likely because of the in-plane tensile strain imposed by the STO substrate. The tensile strain being larger for thinner films. As shown in figure 1 (d), below a critical thickness $t_c$ of ~ 54 nm, a decrease of the out-of-plane lattice parameter is evidenced, while a quasi-plateau observed for thicker films. By decreasing the film thickness below ~ 54 nm, it is therefore possible to continuously tune the tensile misfit-strain felt by the film. Note that the values for the relaxed regime (~3.94 Å) i.e. thick films are close to the bulk pseudo-cubic $b_{pc}$~$b_O/\sqrt{2}$ value ($c_{pc}$~$c_O/2$ = 3.881 Å, $a_{pc}$~$a_O/\sqrt{2}$ = 3.855 Å and $b_{pc}$~$b_O/\sqrt{2}$ = 3.952 Å) [18,19]. This suggests that the NFO thin film growth is along $[110]_O$ orthorhombic direction being parallel to [001]-direction of the STO substrate. To confirm the crystallographic relationships of the NFO thin films with respect to the substrate and investigate the in-plane epitaxial conditions, φ scans for the $(112)_C$ STO and $(132)_O$ NFO family of planes (C and O indices respectively refer to cubic and orthorhombic systems) as shown in Figure 2 for the 212 nm, 160 nm, and 33 nm thick films. Figure 2 indicates a fourfold in plane symmetry of the film relative to the STO crystallographic axis whatever the thickness and confirms that NFO $[110]_O$ axis is along the $[001]_C$ substrate axis. A closer look for thicker films shows orthorhombic twins with a splitting of the diffraction peak in φ-scans while the thinner 33 nm film does not present any splitting and only a single diffraction peak is observed. Reciprocal space mapping around the (103) STO family of planes confirms this structural trend i.e. strain relaxation thanks to elastic domain arrangement, as presented in Figure 2 (g-j). While the 33 nm thin film displays a sharp and single reflection perfectly aligned along $q_z$ (the same $q_x$) with the STO substrate, the 54 nm thick film shows two very close reflections with different $q_x$ values with respect to the STO peak and corresponding to orthorhombic twins. The totally relaxed thick film of 212 nm presents a broad reflection synonymous with some in-plane mosaicity (i.e. epitaxial growth with strong texture). The above results also suggest that a strain-induced orthorhombic to tetragonal-like structure likely takes place in the NFO thin films below the critical thickness $t_c$. Above this thickness, the stress is no more sustained by the film and is released through splitting into orthorhombic domains (Fig S1 provides a sketch of the NFO domains).

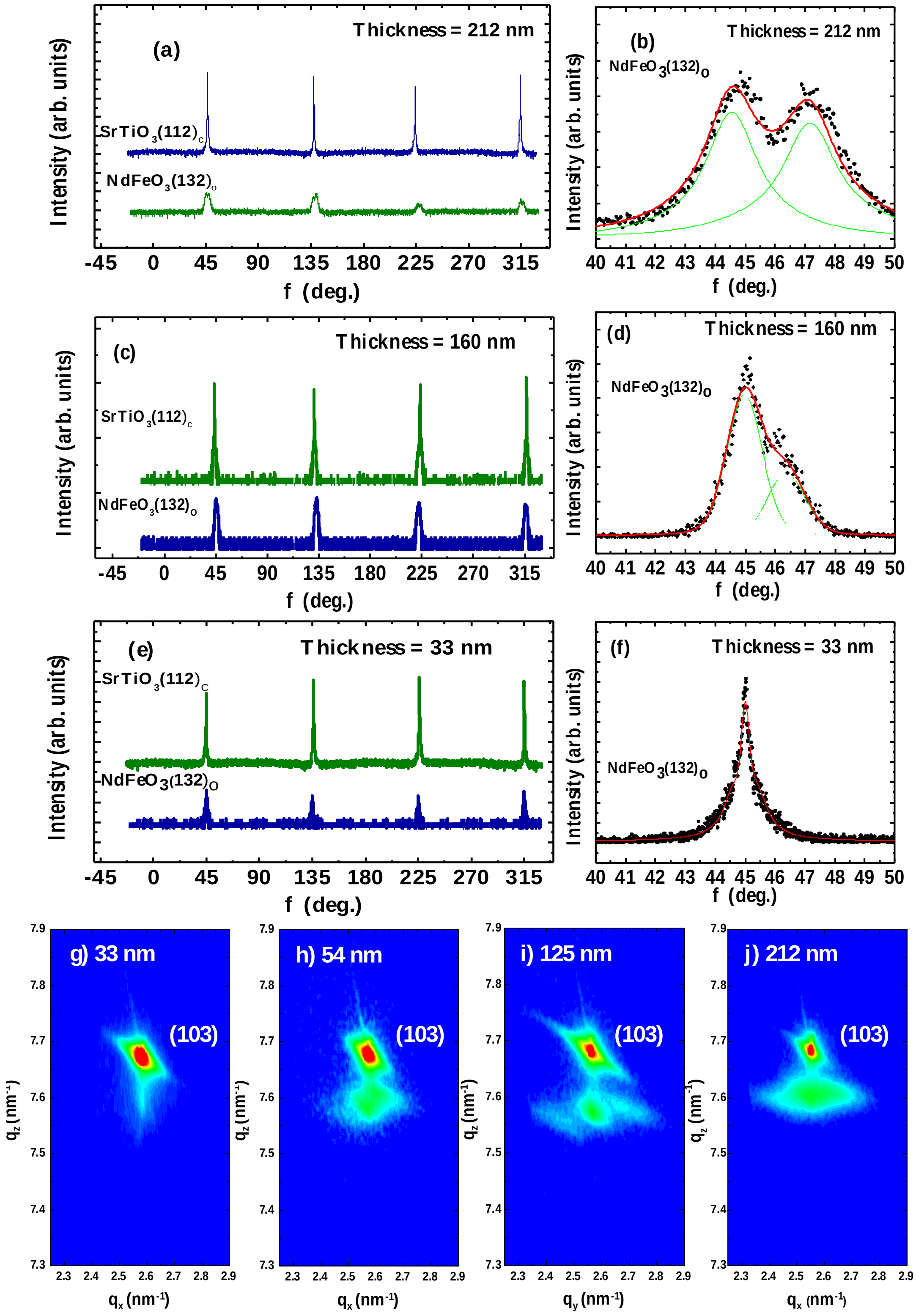

Figure 2. φ scans for the (112)C STO and (132)O NFO family of planes (C and O indices respectively refer to cubic and orthorhombic systems) for (a) 212 nm (c) 160 nm and (e) 33 nm thick films. (b), (d), and (f) correspond respectively to zoomed-in regions at specific angular

range and highlight presence for 212 nm and 160 nm and absence of orthorhombic twinning pattern for 33 nm. Note also the larger width of the doublet for the thicker 212 nm compared to 160 nm which suggests an increase in-plane misorientation. Reciprocal space mapping around the (103) STO family of planes for the (g) 33 nm, (h) 54 nm, (i) 125 nm, and (j) 212 nm thin film.

To directly access to the orthorhombic twinning pattern and confirm the above XRD analysis, High Resolution Transmission Electron Microscopy (HRTEM) have been performed and the images are presented in figure 3 (additional data are shown in the supplementary information in figure S4 and S5 on the orthorhombic twin patterns and vertical domain walls) [10].

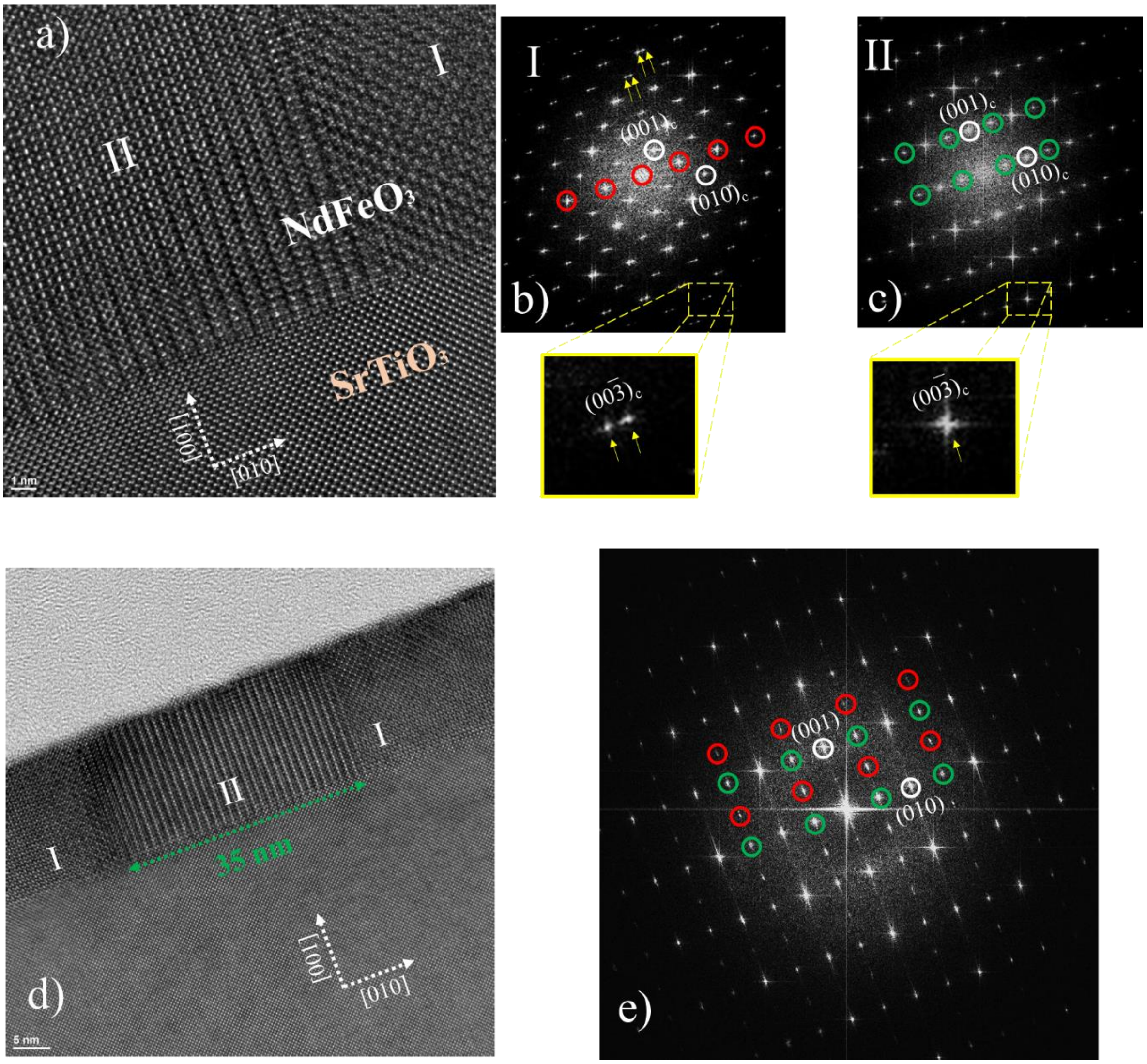

Figure 3. HRTEM analysis of NFO thin films. a) HRTEM cross-section of the 125 nm NFO film with b) and c) Fast Fourier Transform of the two structural domains I and II. In-plane twins

are detected for the orthorhombic domain I (highlighted with two yellow arrows in zoomed-in image). Red circles for domain I correspond to ½(hh0) typical of electron diffraction pattern of Pbnm system with the electron beam parallel to $[001]_O$ orthorhombic crystallographic direction. Doubling of the unit cell is detected for domain II with half order pseudo-cubic diffraction peak ½(0k0) (green circles). d) HRTEM cross-section of the 12.1 nm NFO film with e) FFT of the NFO thin films with both half order diffraction peak (green circles) and ½(hh0) diffraction peaks (coexisting domain I and II). Splitting within domains I are no longer observed.

HRTEM images (Fig. 3) of the NFO thin films show two types of structural domains separated by vertical domain walls (see supplementary information for cross-sections with a larger field of view). The in-plane $c_O$ axis orientation parallel to $[010]_C$ STO axis is confirmed whatever the NFO thickness through the observations of half order diffraction peaks ½(0k0) (see domain II figure 3 and corresponding FFT). The HRTEM investigation indicates zones with $c_O$ axis lying parallel to STO $[100]_C$ (domains I and corresponding ½(hh0) reflections). $\{110\}_O$ NFO family of planes are parallel to the STO surface and define the thin film's growth orientation. Hints about strain relaxation are moreover inferred by observations of splitting of reflections for the thicker NFO films (125 nm) for FFT in domains II only. No splitting and relaxation are detected for the 12.1 nm thick film. These observations are in agreement with the φ scans and reciprocal space mapping (Fig. 2). $c_O$ orthorhombic axis seems to match the in-plane STO lattice parameter (small tensile mismatch of $c_O/2$ of 0.6% with STO and single reflection for domain II) while strain relaxation is observed along projected $[1\text{-}10]_O$ in-plane directions (domains I). This is in agreement with a larger mismatch corresponding to 1.3 % (tensile strain) and -1.2 % (compressive strain) respectively of $a_O/\sqrt{2}$ and $b_O/\sqrt{2}$ with respect to STO parameters (see figure S1 for the different orientations of the NFO unit cell on cubic STO substrates).

The orthorhombic to tetragonal structure induced by tensile strain might be accompanied by a modification of the NFO ($a^-a^-c^+$) octahedra tilt/rotation system (Glazer notation [20-22]). Interestingly Vailionis *et al.* evidenced a similar orthorhombic to tetragonal-like transition for $SrRuO_3$ thin films under tensile strain and an analogy can be made with NFO [23]. Indeed, while bulk $SrRuO_3$ adopts a similar Pbnm system compared to NFO, under tensile strains, the thin films show a tetragonal symmetry with a modified rotation/tilt system that becomes ($a^+a^-c^0$). Such alteration of the oxygen tilt system is also likely to occur in NFO films below the critical thickness $t_c$. Consequently, the superexchange interaction that is driven by Fe-O-Fe angle and Fe-Fe distances should be also impacted and changes of the magnetic properties of the NFO thin films are therefore expected. Confirmation of the in-plane antiferromagnetic ordering is provided by X-ray absorption spectroscopy (XAS) shown in figure 4 for the NFO 23 nm thick film. XAS measured at Fe $L_{2,3}$ edges and for different polarizations (circular versus linear dichroism) is a powerful technique to investigate ferromagnetic and antiferromagnetic ordering. Both total electron (figure 4) and fluorescence yields were measured (figure S6) [10]. The XAS data shown in fig. 4 are typical of $Fe^{3+}$ within an octahedral environment which is an additional proof of parasitic free samples. The weak X-ray magnetic circular dichroism (figure 4 (a)) measured at low temperature with and without an applied 5T magnetic field shows a very weak canted moment typical of orthoferrite systems. The linear dichroism in figure 4 (b) with the line shape similar to $LaFeO_3$ (110) oriented thin films [25], confirms the antiferromagnetic spin ordering and crystallographic orientation. Observation of such linear dichroism strongly suggests a preferential in-plane antiferromagnetic arrangement that is probably connected to the substrate miscut. Only a full angular analysis of the linear dichroism would however identify the precise direction of the Néel AFM vector.

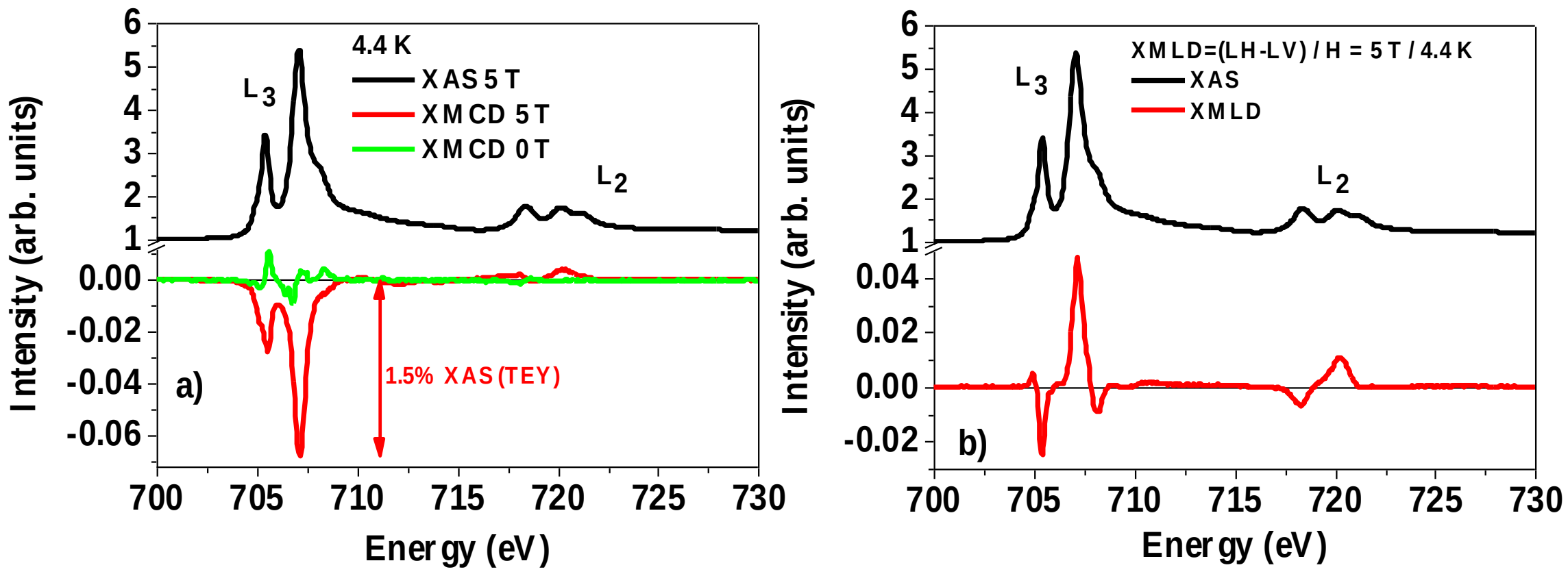

Figure 4. X-ray absorption spectroscopy at Fe $L_{2,3}$ edge in total electron yield (TEY) for a 23 nm thick NFO film at 4.4 K and (a) XMCD: x-ray magnetic circular dichroism (difference between circular right and circular left polarized spectra) at zero and under 5T applied magnetic field. (b) XMLD: x-ray magnetic linear dichroism (difference between linear horizontal and vertical polarized spectra).

In order to gain more information on the magnetic properties, M(H) hysteresis loops were acquired and are shown in Figure 5. Diamagnetic contribution from the STO substrate has been corrected for the M(H) loops analysis.

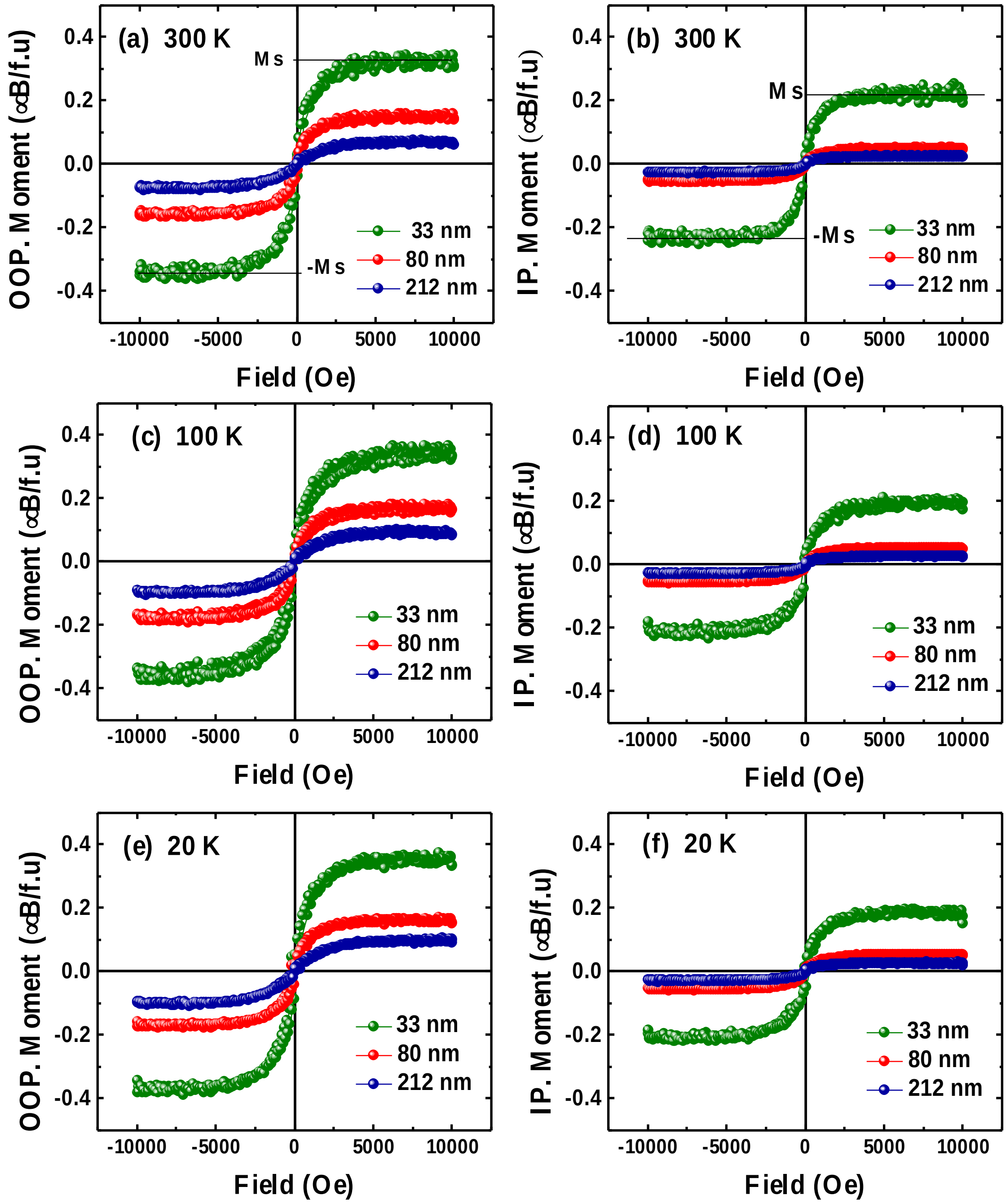

Figure 5. Hysteresis loops measured along the out-of-plane (OOP) and in plane (IP) directions for three different thicknesses and at 300 K (a and b) 100 K (c and d) and 20 K (e and f).

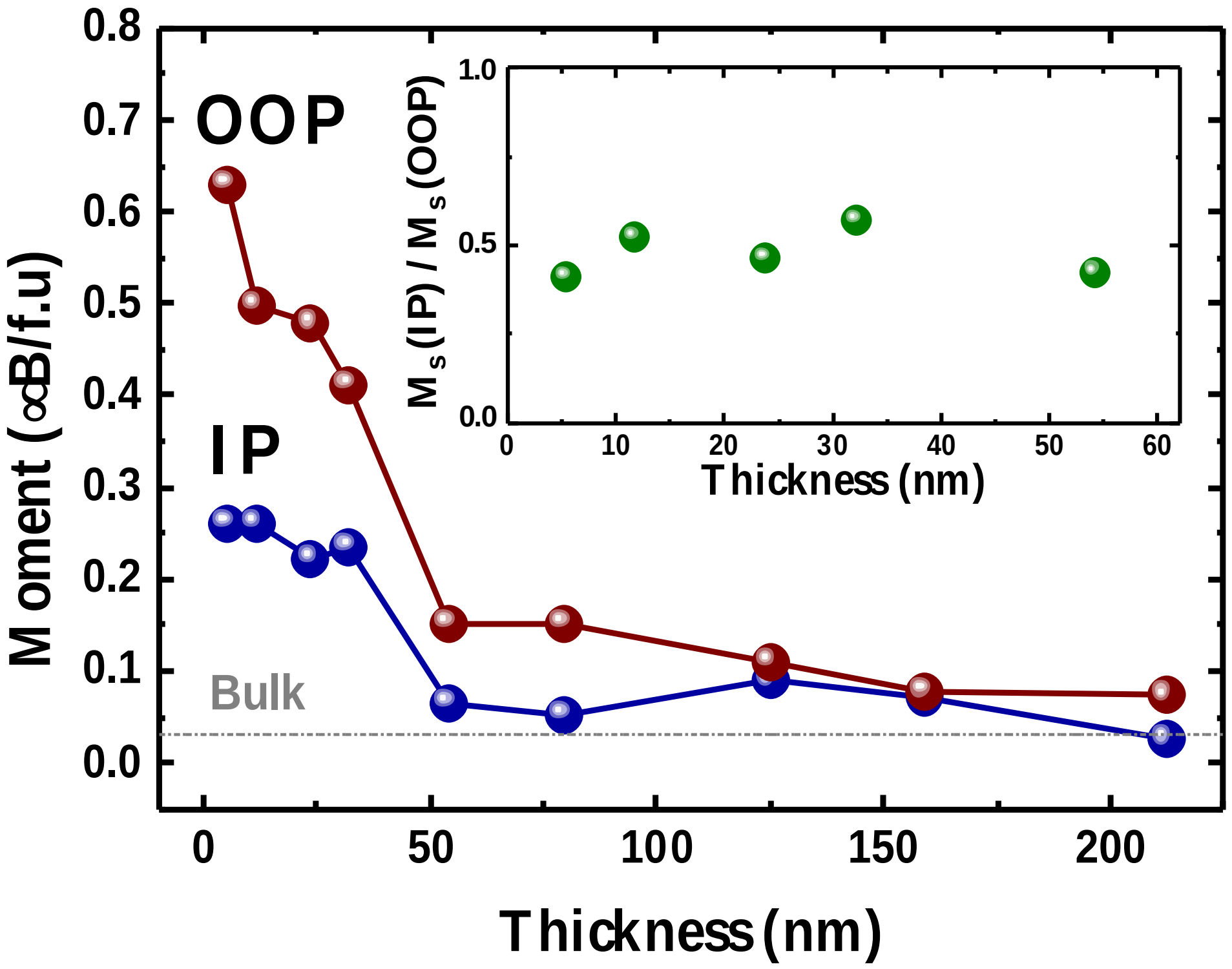

Figure 6. Room temperature in-plane and out of plane Magnetization at saturation (Ms) deduced from the M(H) loops for the whole set of samples. The dashed horizontal line indicates the Ms bulk value. Inset: ratio between in-plane and out-of-plane Ms values versus thickness below the critical 54 nm-thickness.

Very narrow hysteresis loops are observed both in the plane and along the out-of-plane direction for the whole set of samples and over the whole range of temperature investigated (3 K – 300 K). These observations rule out any possible long-range ferromagnetic ordering and are in agreement with an antiferromagnetic state as observed in bulk NFO. A clear trend is observed on the magnetization at saturation (Ms) as a function of thickness and orientation. Indeed, a decrease of Ms is revealed on increasing thickness for both in-plane and the out-of-plane direction with Ms values tending towards bulk ones as highlighted in figure 6. The value of Ms discards from its bulk-like one and its increase becomes significant below the critical thickness of $t_c$ ~ 54 nm where both in-plane and out-of-plane contribution split attesting of the magnetic anisotropy taking place in the thinner films. There is a clear link with the strain release evidenced through appearance of orthorhombic domains we showed using X-ray diffraction and electron microscopy data. Such a correlation is most likely caused by the orthorhombic to tetragonal-like distortion on decreasing thickness and the related domain arrangement. It is also worth mentioning that while the total magnetic anisotropy appears below $t_c$ ~ 54 nm, the ratio of in-plane and out-of-plane contribution of $M_S$ remains roughly constant (see inset fig. 6) indicating that the decrease of film thickness and reciprocally the increase of tensile misfit strain favours equally both components. The increased crystallographic distortion affects the DMI interaction and the weak ferromagnetic spin canting as previously shown for $TmFeO_3$ and $SmFeO_3$ [5,14]. A clear correlation is therefore evidenced for the increased tensile strain below $t_c$ = 54 nm and the concomitant increased ferromagnetic weak canting in both in and out plane directions (probably due to the twinning pattern). Strikingly, in addition to this enhanced magnetic anisotropy, the misfit strain also impacts on the low-temperature magnetic behaviour of the films. While bulk NFO shows a Fe-sublattice spin reorientation transition between 200 K and 100 K and other spin couplings below ~ 30 K related to the presence of Nd-sublattice, different behaviors in the magnetic moment temperature dependence are evidenced for two representative thin films on cooling from 300 K down to 2 K (see figure S7) [10]. Strain-induced modification of the magnetic structure as well as spin reorientation transition have already been observed in other orthoferrites such as $LaFeO_3$ [25] or $DyFeO_3$ [13] as well as in $Fe_2O_3$ [26,27]. Additional investigations (out of plane moment) are however needed to reveal the exact magnetic phase stability and its magnetocrystalline relation to tensile strain induced tetragonal-like state. Further works, especially at high temperature and under magnetic field applied in- and out-of-plane of the films, are requested in order to better characterize the phase transitions. Nevertheless, we clearly evidence how tensile misfit strain, we were able to tune continuously by varying the film thickness, can be used to impacts on the structure (both symmetry and elastic domain configuration) of NFO films enabling to induce a magnetic anisotropy with controlled amplitude of the magnetic moment.

**Conclusions**

Epitaxial orthoferrite NFO thin films were grown by pulsed laser deposition on (001)-oriented $SrTiO_3$ substrates and their structure and magnetic responses were investigated by complementary and advanced tools. XRD and HRTEM investigations indicate excellent crystalline quality with a tensile misfit strain for films with thicknesses below the critical value of 54 nm. While for thick films (above $t_c$ ~ 54 nm), NFO displays a $(110)_O$ growth orientation and strain relaxation via an orthorhombic twinning pattern, thinner films show a tetragonal-like distortion which is continuously enhanced by decreasing the film thickness. As a result, a magnetic anisotropy is induced likely thanks to magnetocrystalline coupling and the amplitude of both in-plane and out-of-plane components can be then continuously enhanced (while their ratio keeps constant) with the increase of the misfit-strain. Furthermore, preliminary low temperature magnetic measurements suggest a modified magnetic phase stability most likely due to the tensile strain induced tetragonal-like strain.

**Acknowledgments**

This work was supported by ANR grant THz-MUFINS (Grant No. ANR-21-CE42-0030), the Conseil Régional des Hauts-de-France, the Fonds Européen de Développement Régional (FEDER), and the European Union Horizon 2020 Research and Innovation actions MSCA-RISE-ENGIMA (No. 778072). The authors would like to thank Cabié Martiane and Thomas Neisius for TEM sample preparation and observation.